# Rapid design space exploration of multi-clock domain MPSoCs with Hybrid Prototyping

Ehsan Saboori
Department of Electrical and Computer Engineering
Concordia University
Montreal, Canada
e_saboo@encs.concordia.ca

Samar Abdi
Department of Electrical and Computer Engineering
Concordia University
Montreal, Canada
samar@encs.concordia.ca

*Abstract*—This paper presents novel techniques of using hybrid prototyping for early power-performance analysis of MPSoC designs with multiple clock domains. The fundamental idea of hybrid prototyping is to simulate a design with multiple cores by creating an emulation kernel in software on top of a single physical instance of the core. However, so far hybrid prototyping has been limited to homogeneous multicores running at the same clock frequency. Moreover, hybrid prototyping has not yet been demonstrated for efficient design space exploration. Our work focuses on enhancing the capabilities of hybrid prototyping, such that it can be applied to realistic multi-clock MPSoC designs as well to perform early power-performance evaluation of MPSoC designs. Our experiments using industrial strength applications such as JPEG, MP3 and Packet Processing, demonstrate the high accuracy of our hybrid prototypes, and over two orders of magnitude improvement over software simulation speed. We also demonstrate that exploring over 150 design options using hybrid prototyping can be done with high reliability in the order of minutes compared to multiple days using conventional FPGA prototyping.

## I. Introduction

Multi-processor System-on-Chip (MPSoC) platforms are becoming increasingly pervasive in modern embedded systems. System-level modeling techniques have enabled creation of fast software models of multicore platforms, commonly known as virtual prototypes, for early functional validation of embedded software, before the hardware is available. Virtual platforms have the advantage of high-speed functional simulation and, typically, scale well with the number of cores. However, they sacrifice timing accuracy for speed. FPGA prototypes provide cycle-accurate performance estimation. However, it takes a significant amount of time to design, implement and test the FPGA design. Hybrid prototyping [1] is a system-level modeling framework that aims to provide early, fast, and cycle-accurate models of homogeneous multi-cores. Using hybrid prototyping, embedded software designers can create concurrent applications and accurately analyze the performance implication of their optimizations before implementation. At the same time, hardware architects can modify the platform model without having to do full FPGA prototyping.

In this paper, we enhance the hybrid prototyping system to simulate MPSoCs with multi-clock domains, which introduce heterogeneity of processing speed amongst the cores. We also introduce methods for early power modeling that enables comprehensive *D*esign *S*pace *E*xploration (DSE) of MPSoC platforms using hybrid prototypes. By using our proposed DSE methods, designers can evaluate their design choices during system development. Hybrid prototyping can be used to implement a set of prototypes for different design choices. The set of prototypes can be compared using well defined metrics such as execution time, cost and energy consumption.

In this paper, our primary goal with hybrid prototyping is to make DSE fast, early and reliable based on execution time and energy consumption metrics for MPSoCs. The contributions of this work are (i) timing estimation with multi-clock domains, (ii) energy consumption estimation and (iii) design space exploration, using hybrid prototyping.

## II. Related work

Both virtual prototyping and FPGA-based physical prototyping have been a topic of intense research with the growing adoption of multicore architectures and the corresponding need to provide early simulation models to embedded software designers. Amongst the software-based methods, the most successful developments have been virtual platform technologies based on binary translation, as commercialized by Windriver [2], Coware [3], and Xilinx XVP [4]. In most virtual platforms, host-compiled ISS have replaced or complemented traditional cycle-accurate micro-architecture simulators [5] [6] [7] [8]. Open Virtual Platform (OVP) is an open source virtual platform which uses OVPsim to simulate different designs. OVPsim is an instruction accurate simulator which provides infrastructure for describing platforms with one or more processors containing shared memory and busses in arbitrary topologies and peripheral models [9]. Such simulators can provide significant speedups (reaching simulation speeds of several hundred MIPS), but often focus on functionality and speed at the expense of limited or no timing accuracy. Another host-compiled software simulation technique is based on source level static delay annotation in the application [10] [11]. The delays are derived by analyzing the application execution on an abstract model of the core. Although source-level annotation techniques promise high simulation speed, they require the full application source, including source of

This work was funded by the *Natural Sciences and Engineering Research Council of Canada* (NSERC) grant no. 386311-2010.

libraries. These techniques also use an abstract core model, leading to estimation inaccuracies.

The RAMP platform puts together a large array of FPGAs in order to support the instantiation and integration of hundreds of cores [12] [13]. Unfortunately, the cost and design time of such full system prototypes is very high [14]. In addition, there is no flexibility of abstracting the inter-core communication in RAMP, since it is fixed in hardware by the inter-FPGA communication architecture. Another FPGA-based modeling approach implements the SystemC simulation kernel in FPGA to support standard hardware I/O during simulation [15]. Yet another type of FPGA-assisted simulation, called virtual in-circuit emulation, runs software-on-host and application-specific hardware on FPGA to avoid slow RTL simulation in software [16]. Both techniques are incremental improvements to cycle-accurate simulation and have not been shown to scale to large multicore designs.

Hybrid prototyping technique is distinct from the above approaches in that it time-multiplexes several virtual cores on a single physical target core. The core and the additional simulation infrastructure can fit on a single FPGA chip, making it very cost effective in contrast to full system prototyping in FPGA. The MEK supports the execution of any multi-tasking ANSI C/C++ application in our proposed environment.

### III. BACKGROUND ON HYBRID PROTOTYPING

The fundamental idea of hybrid prototyping is to create a multicore emulation kernel (MEK) in software that executes on a single target core that is physically implemented in FPGA [1]. Figure 1 shows the MEK structure in grey. The lowest layer is the MEK data structure, consisting of tasks, events and the scheduling First-In-First-Out (FIFO) queue that keeps all the ready tasks. The next higher layer consists of the simulation primitives for the management of events and logical times for the tasks.

Fig 1. The layered MEK structure for hybrid prototyping

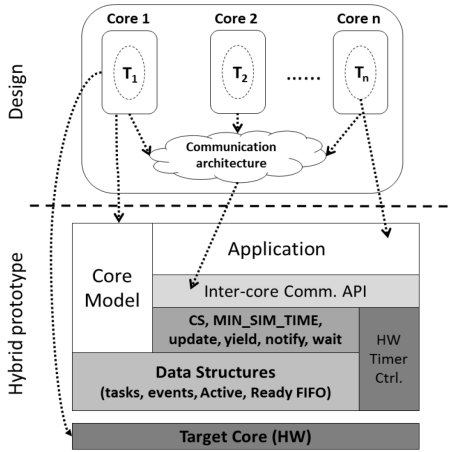

The models of the inter-core communication services are implemented as an API on top of the simulation primitives. The architecture is modeled using task creation and channel creation methods as shown. The application consists of the task functions and uses the communication API provided by the MEK. Since, hybrid prototyping depends on measurement of physical time, a hardware timer and its drivers are included.

The MEK defines primitives for event management using basic notify/wait methods. It also defines an update method to manage the logical times of the emulated cores. The logical time for a core is the time until which the core has been simulated. The context switching between emulated cores during simulation is done by the context_switch method that saves the context (program, stack pointers and registers) of the previous core and loads the context of the new Active core. A running core may yield the simulation priority by calling the yield method. Finally, the MEK maintains a global variable called MIN_SIM_TIME, which keeps the minimum logical time until which all cores have been simulated.

#### A. Timing estimation

Since, hybrid prototyping depends on the measurement of physical time, a hardware timer and its drivers are included. The MEK uses a hardware timer to measure the execution time in CPU cycles. The timer's value is used to manage the emulated core's logical time. It can be measured either in CPU cycles or real execution time (in milliseconds). To obtain execution time in millisecond, a clock frequency ($f$) should be assigned to each emulated core. As the MEK calculates the execution time in CPU cycles, the real execution time can be easily obtained by multiplying the number of cycles with the clock period of each emulated core.

$$execution\ time_{ms} = \text{CPU Cycles} \times \frac{1}{emulated\ core's\ frequency}$$

Figure 2 shows how the MEK maintains the logical times $lt_1$ and $lt_2$ on emulated cores $C_1$ and $C_2$ respectively. It is assumed that the length of the channel is only 1 item and $C_1$ and $C_2$ are running with frequencies $f_1$ and $f_2$ respectively.

The MEK may pick either $C_1$ or $C_2$ to simulate first. Figure 2 illustrates the case when $C_1$ is picket by the MEK. The X-axis shows the physical time measured by the hardware timer in CPU cycles. As this figure shows, the MEK runs $T_1$ on emulated core $C_1$ until it reaches bread at time $t_{11}$. Since no data is found in the channel, the MEK update the $lt_1$ to $t_{11} \times f_1$, blocks $T_1$ and switches to task $T_2$. $T_2$ runs for $t_{21}$ unit, writes data into the channel then notifies event ch→ev_write. As a result, the MEK unblocks $T_1$. Right after this notification, $T_2$ executes for another $t_{22}$ unit. Then, it tends to write for the second time. However, as there is no space in the channel, the MEK blocks $T_2$, set the $lt_2$ to $(t_{21} + t_{22}) \times f_2$ and switches back to $T_1$. $T_1$ reads data from the channel and notifies event ch→ev_read. Considering this notification, the MEK unblocks $T_2$. As $T_1$ can execute for $t_{12}$ unit till reaches the second read while the channel is empty, the MEK sets $lt_2$ to $(t_{11} + t_{12}) \times f_1$ and switches to $T_2$. When $T_2$ writes into the channel, the MEK updates the $lt_1$ to $lt_2$'s value. This happens because $lt_2$ is bigger than $lt_1$. $T_2$, then, executes for $t_{23}$ unit and terminates. The MEK switches back to $T_1$. It reads from the channels, executes for $t_{13}$ units and then terminates. At the end of the simulation, the MEK reports $lt_1$ as $(t_{21} + t_{22}) \times f_2 + t_{13} \times f_1$ and $lt_2$ as $(t_{21} + t_{22} + t_{23}) \times f_2$.

Fig 2. The MEK timing estimation example with two running tasks

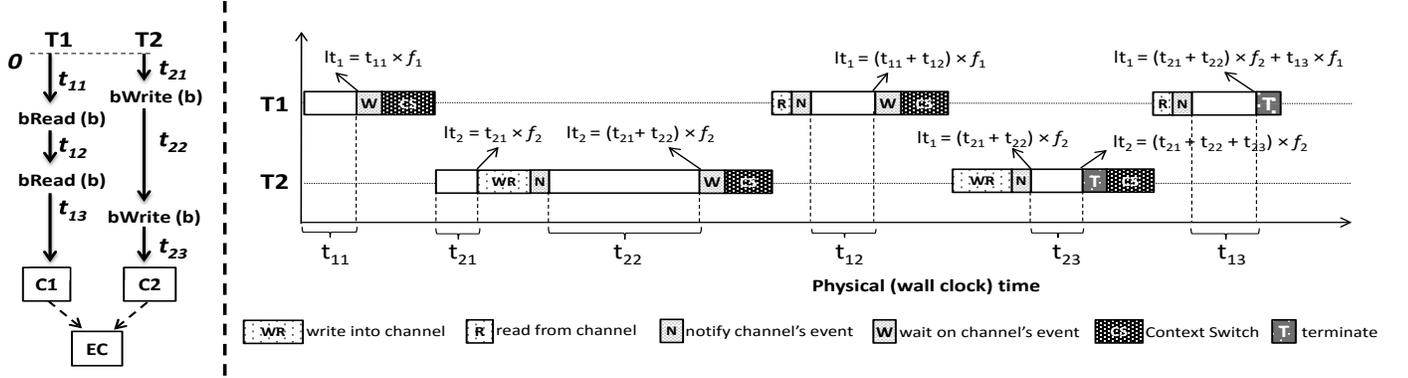

## IV. ENERGY ESTIMATION

The MEK provides the busy and idle time for each cores which are used for energy consumption estimation. On every logical time update after blocking, the difference between the new logical time and the task's logical time indicates the idle time for the corresponding core. Figure 3 shows the busy time for tasks $T_1$ and $T_2$. In this case, when $T_1$ notifies event $e$, the MEK updates $T_2$'s timestamp to $t_{11}$ and increases $T_2$'s idle time by $t_{11} - t_{21}$.

Fig 3. The busy time and idle time

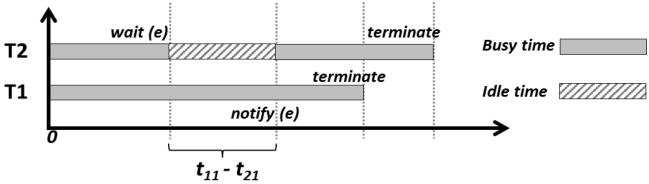

Energy consumption is one of the most important quality metrics in embedded system design. The power consumption of a core is directly related to its frequency. Most embedded processors support several operating frequencies, which allows us to create a mixture of cores, each running at a different operating point. The busy power consumption is a measure of the power which is consumed by the core when it executes the instructions. The idle power consumption is a measure of the, largely static, power consumed by the core while it waits on external events, and does not execute any instruction. We used the Xilinx XPower analyzer [4] to measure both the busy and idle power consumptions. The idle dynamic power is zero for the Microblaze when it waits on the FSL communication channels. The Static power, consumed at all times irrespective of whether the core is busy or idle, is the same for all cores with different clock domains and is measured to be 1.48 mw. If the clock frequency is increased, the power consumption will increase as well. As CPU and memory are the most power consuming parts in our designs, we consider the busy power as sum of CPU power and memory power. Table I shows the average busy power for Microblaze and BRAM.

A simplistic, yet reasonably accurate, power model of a processor assigns a single power consumption number to each operating point. Clearly, the processor is only consuming dynamic power when it is busy. Since different mappings may result in different busy times for the cores, we can change the mapping in order to obtain the best energy consumption by the design. Using a hybrid prototype, the designer can quickly obtain the busy times for the design with different operating frequencies and mappings. The estimated energy consumption for each emulated core can be calculated by the following equations.

$$\text{Energy} = (\text{Idle}_{time} \times \text{Idle}_{power}) + (\text{busy}_{time} \times \text{busy}_{power})$$

$$\text{busy}_{power} = \text{CPU}_{busy_{power}} + \text{Memory}_{busy_{power}}$$

TABLE I. THE BUSY POWER CONSUMPTION FOR DIFFERENT CLOCK DOMAINS

| Frequency | Microblaze | BRAM |
|---|---|---|
| 25 MHz | 07.14 mw | 14.57 mw |
| 45 MHz | 12.23 mw | 25.68 mw |
| 55 MHz | 14.65 mw | 30.80 mw |
| 60 MHz | 16.00 mw | 34.01 mw |
| 90 MHz | 23.24 mw | 50.65 mw |
| 125 MHz | 31.91 mw | 68.91 mw |

## V. USE CASES

To evaluate the speed and accuracy of hybrid prototypes, we used the JPEG encoder, the MP3 decoder and a simple packet forwarding applications. We chose the Microblaze core from Xilinx for the target multicore architectures [4]. The FIFO communication between the tasks is performed using the Fast Simplex Link (FSL) buses supported by Microblaze.

### A. MP3 Decoder

The MP3 decoder application reads and decodes data from the media file. The MP3 data is fetched from a file, and after being decoded it is written into a serial buffer. The buffered data can be played on the handset speaker. This application has 5 separate tasks which can be run on different cores: isrPulser which is responsible for sending pulse in proper time to task isr. Task isr is the interrupt handler that notifies the decoding task if more data is needed by the serial buffer for the speakers. Task audiosal which reads and decodes data from the media file. Task mixerctrl is in charge of the channel and task dspaudio converts the rate, playback, mix, etc. on the data. Figure 4 shows the MP3 decoder application.

Fig 4. The MP3 decoder application

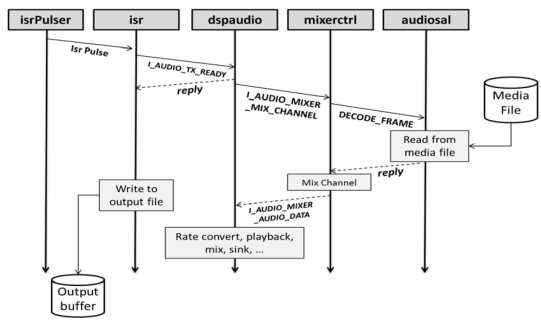

### B. Jpeg Encoder

The JPEG encoder consists of 5 tasks: Read the bitmap (Read), Discrete Cosine Transform (DCT), Quantization of values (Quant), ZigZag transform (ZigZag) and Huffman encoding (Huff). Each function consumes a frame, which is an 8×8 block of integers, processes it and passes the block to the next function. Given the application structure, it can be easily pipelined and the concurrent tasks can be mapped to different cores. We created 15 multicore designs of the JPEG encoder, ranging from 2 core to 5 cores. For each platform, we used different mappings from tasks to cores as shown in Table II.

TABLE II. TASK MAPPINGS FOR THE JPEG ENCODER MULTICORE DESIGNS

| Design | #Cores | Mapping |
|---|---|---|
| 2a | 2 | Read → mb1; DCT, Quant, ZigZag, Huff → mb2 |
| 2b | 2 | Read, DCT → mb1; Quant, ZigZag, Huff → mb2 |
| 2c | 2 | Read, DCT, Quant → mb1; ZigZag, Huff → mb2 |
| 2d | 2 | Read, DCT, Quant, ZigZag → mb1; Huff → mb2 |
| 3a | 3 | Read → mb1; DCT → mb2, Quant, ZigZag, Huff → mb3 |
| 3b | 3 | Read → mb1; DCT, Quant → mb2; ZigZag, Huff → mb3 |
| 3c | 3 | Read → mb1; DCT, Quant, ZigZag → mb2; Huff → mb3 |
| 3d | 3 | Read, DCT → mb1; Quant, ZigZag → mb2; Huff → mb3 |
| 3e | 3 | Read, DCT → mb1; Quant → mb2; ZigZag, Huff → mb3 |
| 3f | 3 | Read, DCT, Quant → mb1; ZigZag → mb2; Huff → mb3 |
| 4a | 4 | Read → mb1; DCT → mb2; Quant → mb3; ZigZag, Huff → mb4 |
| 4b | 4 | Read → mb1; DCT → mb2; Quant, ZigZag → mb3; Huff → mb4 |
| 4c | 4 | Read → mb1; DCT, Quant → mb2; ZigZag → mb3; Huff → mb4 |
| 4d | 4 | Read, Quant → mb1; DCT → mb2; ZigZag → mb3; Huff → mb4 |
| 5 | 5 | Read → mb1; Quant → mb2; DCT → mb3; ZigZag → mb4; Huff → mb5 |

### C. Packet forwarding application

The JPEG encoder and the MP3 decoder can be run on a design with maximum number of 5 cores. However, to evaluate the overhead of the hybrid prototyping and also to show its scalability, we need an application that can be run on a large number of cores simultaneously. A packet forwarding application would be an ideal choice for this purpose. Therefore, a simple application has been implemented in order to process packets. The application has a dispatcher responsible for reading packets and distributing them among the inner-cores. The inner-cores execute packet processing tasks and send the processed packets to the collector. The collector receives all packets and puts them in a proper order. This application can be implemented with a large number of inner-cores that can each be implemented as a Microblaze in FPGA prototype. The cores are connected using FSL as shown in Figure 5.

Fig 5. Simple Packet forwarding application

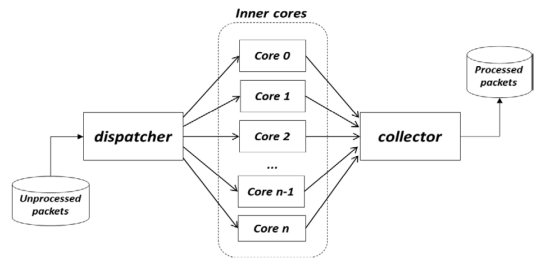

## VI. EXPERIMENTAL RESULTS

We created a FPGA prototype, a hybrid prototype and a virtual prototype for different designs for JPEG encoder, MP3 decoder and packet forwarding applications. All the used Microblaze cores are clocked at 125 MHz. Each Microblaze core in the FPGA prototypes has 64 KB of dedicated Block RAM (BRAM) for program and data. The hybrid prototypes use a single Microblaze core with 64 KB of BRAM since all the tasks and the MEK fit in a single BRAM. For larger programs, one may create multiple instances of BRAMs with contiguous address space assignment. OVP is used to create the virtual prototypes. As OVP is an instruction accurate simulator, it only calculates the number of instructions and cannot measure the idle time. Therefore, the busy time for each core is the sole result that can be provided by the OVP.

### A. Speed

Figure 6 shows speed comparison between hybrid, FPGA and virtual prototypes to execute the JPEG encoder for a given image. The X-axis is the number of cores and the Y-axis is the simulation time in milliseconds. The real execution time can easily be obtained by multiplying the number of cycles with the clock period of 8 ns (125 MHz).

Fig 6. Prototyping speed comparison between FPGA, hybrid and OVP prototypes for the JPEG encoder

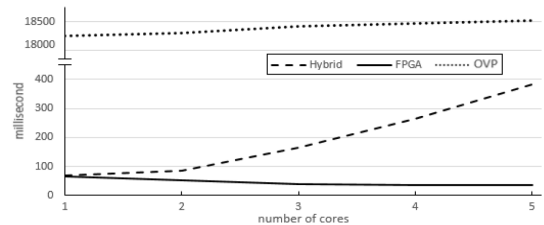

As we can see, in the most complex design with 5 cores, the hybrid prototype took 382 ms to simulate the JPEG encoder. On the other hand, the FPGA prototype took 35. In contrast, the virtual prototyping using OVP took over 18 seconds on a 2GHz Pentium host with 8GB of RAM.

Figure 7 shows the speed comparison between hybrid, FPGA and OVP prototypes for packet forwarding application. The X-axis is the number of cores used in each design and the Y-axis is the simulation time in milliseconds. Up to eight Microblazes can be used on the FPGA due to Microblaze Debug Module (MDM) restriction. MDM can be connected to the maximum of eight Microblazes at the same time. Therefore, FPGA prototypes can be implemented with only up to eight cores. As it shows, in the design with eight cores, the FPGA prototype took 11 ms to execute, while the hybrid prototypes took 131 ms to emulate the application. In contrast, the OVP took 26 seconds to simulate the design.

Fig 7. Prototyping speed comparison between FPGA, hybrid and OVP prototypes for Packet forwarding application

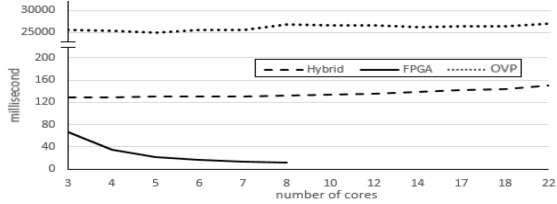

The FPGA prototype took 33M cycles (264 ms) to execute the MP3 decoder while the hybrid prototype took 47.25M cycles (378 ms) to emulate it and OVP took about 28 second to simulate the design. Figure 6 and 7 show that the hybrid prototype emulation time increases linearly when the number of cores are being increased.

### B. Accuracy

Figure 8 shows the busy time reported by FPGA, Hybrid and virtual prototypes for each core for all 15 different designs mentioned in table II. The X-axis shows the designs and Y-axis shows the execution time in million cycles for each design.

Fig 8. The busy times for FPGA, hybrid and Virtual prototypes for the JPEG encoder

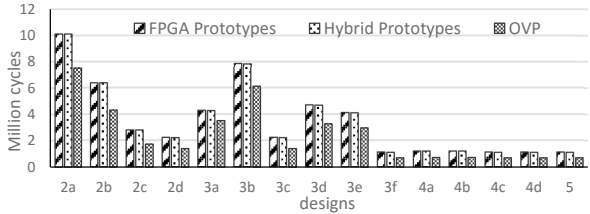

Both FPGA and hybrid prototypes were created for packet forwarding application. As it was mentioned earlier, up to eight Microblazes can be used on the FPGA due to MDM restriction. However, as there is no such limitation in hybrid prototypes, as they use a single Microblaze, the hybrid prototypes can be easily implemented for designs with more than 8 cores. Figure 9 illustrates the results for designs with up to 8 cores. The X-axis shows the number of cores and Y-axis shows the execution time in million cycles for each design.

Fig 9. Packet forwarding application execution time for all designs with up to 8 cores

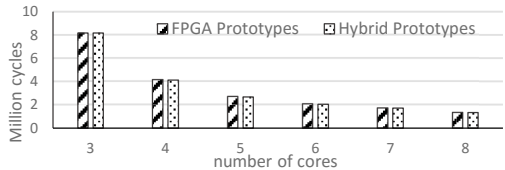

The MP3 decoder has only one design with 5 cores. Table III shows the results for FPGA, hybrid and virtual prototypes of the MP3 decoder.

TABLE III. THE MP3 DECODER EXECUTION TIME IN CPU CYCLES

| Core | execution time | | Accuracy | Busy time ratio | | |
|---|---|---|---|---|---|---|
| | FPGA | Hybrid | | FPGA | Hybrid | OVP |
| 1 | 31246757[1] | 31250217 | 99.98 % | 100% | 100% | 96% |
| 2 | 32890978 | 32719520 | 99.47 % | 22% | 23% | 0% |
| 3 | 33058711 | 32908133 | 99.54 % | 28% | 28% | 26% |
| 4 | 32972069 | 32839963 | 99.59 % | 17% | 17% | 5% |
| 5 | 32962442 | 32780419 | 99.44 % | 14% | 14% | 45% |

To evaluate hybrid prototypes with multiple clock domains we also created both hybrid and FPGA prototypes for the MP3 decoder and the JPEG encoder by running each emulated core with different clock frequencies (60, 90, 25, 45 and 55 MHz). Table IV contains the results for the MP3 decoder.

TABLE IV. MP3 DECODER RESULTS WITH MULTIPLE CLOCK DOMAINS

| Core | Clock | execution time | | Accuracy | Busy time ratio | |
|---|---|---|---|---|---|---|
| | | FPGA | Hybrid | | FPGA | Hybrid |
| 1 | 60 MHz | 518.69[a] | 520.83 | 99.58 % | 100% | 100% |
| 2 | 90 MHz | 559.1 | 560.8 | 99.69 % | 14% | 14% |
| 3 | 25 MHz | 560.7 | 566.0 | 99.06 % | 43% | 43% |
| 4 | 45 MHz | 560.1 | 563.3 | 99.43 % | 22% | 22% |
| 5 | 55 MHz | 560.0 | 561.9 | 99.66 % | 15% | 15% |

[a.] The time unit is milliseconds.

Figure 10 shows the results for all 15 designs mentioned in table II for the JPEG encoder. The X-axis shows the designs and Y-axis shows the execution time in million cycles for each design.

Fig 10. The execution times for FPGA and hybrid prototypes for the JPEG encoder with multiple clock domains

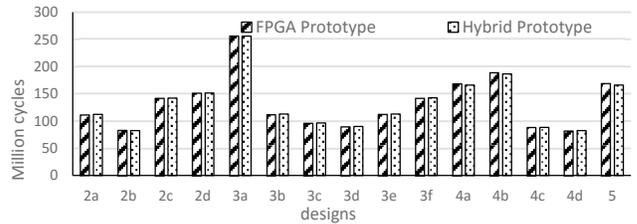

As it was described earlier, all the inner cores in packet forwarding application are doing same processing, therefore, there would be no point to use multiple clock domains for it.

Based on the experimental results, the hybrid prototypes reported the same number of cycles for each task as measured by the FPGA prototypes. This is because the hybrid prototype executes the tasks on the same core as in the FPGA prototype. In contrast, because of the high abstraction level of the underlying ISS, OVP simulation had an error of over 25% in the number of cycles reported. Furthermore, OVP can only report busy time for each core because it is an instruction accurate simulator. Therefore, hybrid prototype was shown to be more reliable than abstract virtual prototypes.

### C. Modeling Effort

Modeling effort is a difficult metric to measure because of the human element. In creating our experimental setup, we found it was very difficult to debug the FPGA prototypes as the number of cores increased. We used a JTAG based debug module provided in the Xilinx Embedded Development Kit.

The I/O from the different cores was sent to the hyper-terminal on the host. In the case of multiple cores, it was difficult to sort through the debug messages from the different cores. In the hybrid prototype, we had to interface with only one core, and the state of the core being emulated was easily observed at any given time. In summary, we found it much more challenging to implement and validate the FPGA prototypes than the hybrid ones.

## VII. Design space exploration

For a given application and architecture, there are several possible mappings. Different mappings are created and evaluated as per the chosen quality metrics. Eventually, the designer can select the best design amongst the evaluated mappings. We consider execution time and energy consumption as quality metrics for DSE.

For each design, the hybrid prototype provides a simple energy consumption model and a highly accurate estimation of the application's execution time. The timing estimations are generated for both total execution time and busy time for each core. In a given multicore design, the longest execution time amongst all tasks (mapped to different cores) can be considered as the design's total execution time. So by comparing the total task execution times on all cores, we can determine the speed of a multi-core design.

Fig 11. Speed vs. Energy consumption for 162 possible designs for the JPEG encoder with 2 clock domains

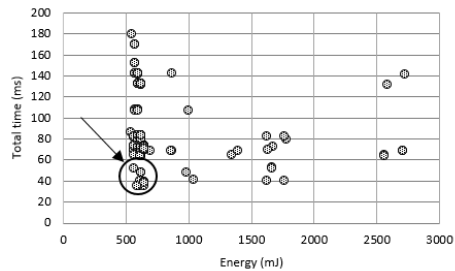

Our experiments with the hybrid prototyping system demonstrate its applicability to fast multicore design space exploration. Embedded system designers have several design choices based on the number of cores, their configuration, and the mapping of the application tasks to those cores. For instance, there are 16 possible designs with only 1 clock domain while there are 14406 different possible designs with 6 clock domains. By increasing the number of clock domains, the number of possible designs increases dramatically. Therefore, it is impractical to implement all these designs and choose the best one. We have modeled the 162 possible designs of the JPEG encoder which is being run with 2 clock domains (60 and 125 MHz). Xilinx Virtual Platform (XVP) simulation shows errors of over 40% in the number of cycles reported because of its high abstraction level. The FPGA prototype takes 35ms to execute and 15 minutes to synthesize every design choice. Therefore, full FPGA prototyping takes almost 40 hours for all 162 possible designs without considering the effort of creating the FPGA prototypes. In contrast to the above techniques, it takes only 15 minutes to synthesize the hybrid platform's target core, which is a one-time effort. The hybrid prototype takes 382ms (in worst case) to emulate each design, thereby enabling extremely fast, early and reliable design space exploration. Figure 11 plots speed vs. energy consumption reported by the hybrid prototypes for all 162 designs which each spot presents a design. The circle highlights the best designs that consume minimal energy and shortest execution time.

## VIII. Conclusion

In this paper we have presented enhancement to hybrid prototyping such that it can applied to design space exploration of realistic multi-clock domain MPSoCs. Using hybrid prototypes, MPSoC designers can create concurrent applications and accurately analyze the power and performance implication of their optimizations before the hardware is available. As such, the hybrid prototyping was proven capable of fast and early MPSoC design space exploration. MPSoC architects can optimize the hardware architecture without having to do full system prototyping. In the future, we will extend the hybrid prototyping approach to support MPSoCs comprised of processors with different instruction-set architectures in order to support fully heterogeneous MPSoC design.